\documentstyle[preprint,aps]{revtex}
\tightenlines
\preprint{\vbox{  
\hbox{IFT-P.019/98}   
\hbox{hep-ph/9803450}
\hbox{March 1998} }}  
\begin{document}  
\draft 
\title{
Left-right asymmetries in polarized $e$-$\mu$ scattering}
\author{J. C. Montero, V. Pleitez and M. C. Rodriguez} 
\address{
Instituto de F\'\i sica  Te\'orica\\ 
Universidade  Estadual Paulista\\
Rua Pamplona, 145\\ 
01405-900-- S\~ao Paulo, SP\\ 
Brazil } 
\date{today}
\maketitle 
\begin{abstract}  
We consider, in the electroweak standard model context, several left-right 
asymmetries in $\mu e$ elastic scattering at fixed target and collider 
experiments. For the former case, we show that the muon mass effects are 
important in a wide energy range. We also show that these asymmetries
are sensitive to the electroweak mixing angle $\theta_W$. The effect of
an extra $Z'$ neutral vector boson appearing in a 331 model is also considered.
The capabilities of these asymmetries in the search of this extra $Z'$
are addressed.
\end{abstract}
\narrowtext   
  
%\section{Introduction}
%\label{sec:intro}  
Several years ago Derman and Marciano~\cite{dm} considered briefly the 
left-right asymmetry in $\mu^-e^-\to \mu^-e^-$, 
here denoted by $A_{RL}(\mu e)$, with unpolarized muon scattering 
on polarized electron target. According to them this asymmetry can be measured 
and the only difference between the high energy $\mu^-e^-$ scattering and the 
high energy $e^-e^-$ scattering is that in the former we must omit crossed 
amplitudes. With the attention received by muon colliders nowadays this issue 
become important~\cite{gunion}. Beside, it has been shown that the mass effect 
of the muon can not be neglected in muon-muon scattering~\cite{assi1}. 
Here we will show that at the Born approximation and in the context of
the electroweak standard model there are appreciable effects coming from the
mass of the muon.

Some advantage to measure $A_{RL}(\mu e)$ were commented
in Ref.~\cite{dm}. The first is that muon beams generally have energies at 
least one order of magnitude greater than the electron beams. Thus
experimentally, since the left-right asymmetry in $ee$ and in $\mu e$ are
proportional to $s$, a larger value for the asymmetry in the later case
can be obtained. Secondly, in $\mu^- e^-$ scattering the background contribution
from $\mu^-N$ scattering is less severe because one could trigger on a single 
scattered electron which could not have arisen from a $\mu^- N$ collision.
The problem arisen by Derman and Marciano was that the experiment requires
strictly zero muon polarization, otherwise the purely QED spin-spin effect
would totally mask $A_{RL}(\mu e)$. However we will show below that this
happens only for low values of $y \equiv \sin^2(\theta/2)$.
This problem does not exist in a muon-electron fixed target experiment using
the high-energy muon beam M2 of the CERN SPS as in the NA47 
experiment~\cite{smc}. In this case left- and right-handed polarized muon can 
be scattered off unpolarized electron (see below). 

Here we will calculate the $A_{RL}(\mu e)$ asymmetry taking into account
the muon mass but neglecting the electron mass. Recently, ${\cal O}(\alpha^3)$  
purely  QED  spin-spin effect calculations (with both particles polarized) 
have been done in these conditions~\cite{rusos2}. The fact is that 
at NA47 typical energies the muon energy is $E_\mu=190$ GeV, {\it i. e.}, 
the invariant $s$ is only 20 times
bigger than the muon mass squared and the effects of the muon mass cannot be 
neglected. Hence, future electroweak radiative corrections to the
$A_{RL}(\mu e)$ asymmetry probably will have to take into account the muon mass
also.

The left-right asymmetry with one of the particle being unpolarized is defined 
as
\begin{equation}
A_{RL}(\mu e \to \mu e)\equiv A_{RL}(\mu e)=
\frac{d\sigma_R-d\sigma_L}{d\sigma_R+d\sigma_L},
\label{asy1}
\end{equation}
where $d\sigma_{R(L)}$ is the differential cross section for one right
(left)-handed lepton $l$ scattering on an unpolarized lepton $l'$. 
That is
\begin{equation}
 A_{RL}(\mu e)=
\frac{(d\sigma_{RR}+d\sigma_{RL})-(d\sigma_{LL}
+d\sigma_{LR})}{(d\sigma_{RR}+d\sigma_{RL})+(d\sigma_{LL}
+d\sigma_{LR})},
\label{asy2}
\end{equation}
where $d\sigma_{ij}$ denotes the cross section for incoming leptons with
helicity $i$ and $j$, respectively, and they are given by 
\begin{equation}
d\sigma_{ij}
\propto\sum_{kl}\vert M_{ij;kl}\vert^2,\quad i,j;k,l=L,R.
\label{dsigma}
\end{equation}
Notice that
since the $\mu e$ scattering is non-diagonal $d\sigma_{RL}\not=d\sigma_{LR}$, so
the asymmetry in Eq.~(\ref{asy2}) is different from the asymmetry defined as
$(d\sigma_{RR}-d\sigma_{LL})/(d\sigma_{RR}+d\sigma_{LL})$.

We will consider the process
\begin{equation}
e^-(p_1,\lambda)+\mu^-(q_1,\Lambda)\to e^-(p_2,\lambda')+\mu^-(q_2,\Lambda'),
\label{eumu}
\end{equation}
where $q=p_2-p_1=q_2-q_1$ is the transferred momentum. As we said before, 
we will neglect the electron mass but not the muon mass {\it i.e.}, 
$E=\vert\vec{p}_e\vert$ for the electron and 
$K^2-\vert\vec{q}_\mu\vert^2=m_\mu^2$ for the muon. In the non-diagonal 
elastic scattering in the standard model we have only the $t$-channel 
contribution. The relevant amplitudes are in Appendix~\ref{sec:aa} and 
in the appendices of Ref.~\cite{assi1}.

%We will study here $A_{RL}(\mu e)$ in the context of the electroweak 
%standard model (ESM) and in a model with doubly charged bilepton fields, two 
%scalars and one vector (the latter one is denoted by $U^{--}_\mu$) and an 
% extra 
%neutral vector boson $Z'$~\cite{pp}. We identify each case this by using the 
%$\{{\rm ESM}\}$, $\{{\rm ESM+Z'}\}$ and $\{{\rm ESM+U}\}$ labels in 
%cross sections and asymmetries. The doubly charged scalar bilepton 
%contributions cancel out in the numerator of the $A_{RL}$
%asymmetry and we will not consider it. 

We will study here $A_{RL}(\mu e)$ defined in Eq.~(\ref{asy2}) in the context 
of two models: the electroweak standard model (ESM) and in a model having 
a doubly charged bilepton vector field ($U^{--}_\mu$) and an extra neutral 
vector boson $Z'$~\cite{pp}. 
The latter model has also two doubly charged scalar bileptons but since their 
contributions cancel out in the numerator of the asymmetry we will not 
consider them. We identify the case under study by using the $\{{\rm ESM}\}$, 
$\{{\rm ESM+U}\}$, and $\{{\rm ESM+Z'}\}$ labels in cross sections 
and asymmetries.

First, we will consider the $\mu e$ elastic scattering in the context of the
electroweak standard model.
The $A_{RL}$ asymmetry for $m_\mu=0$ and fixed target experiments is
\begin{equation}
\left.A_{RL}^{FT;ESM}(\mu e)\right\vert_{m_\mu=0}=4\beta_Wg_Vg_A\frac{ys}{1+(1-y)^2};
\quad \beta_W\equiv \frac{G_F}{\sqrt{2}} \frac{8 M_W^2}{4\pi \alpha M_Z^2};
\quad y\equiv \sin^2(\theta/2),
\label{asydm}
\end{equation}
which is the expression obtained by Derman and Marciano~\cite{dm}. 
Here we will not shown explicitly the $A_{RL}$ expression for 
$m_\mu\not=0$. For
$y=1/2$, the above asymmetry is $A_{RL}^{FT}(\mu e)=-1.643\times10^{-6}\,s$.
Although it is a rather small value the cross section still depends on 
$m_e^{-1}$ implying a large cross section which can allow to have enough 
statistics as in the case of $ee$ scattering~\cite{dm}. In fact,
from Eq.~(\ref{ab1})  we obtain a total cross section of $11.86(3.88)$ mb for 
$E_\mu=50(190)$ GeV. (In Appendix~\ref{sec:ab} we shown
the differential cross section for $\mu e$ scattering in the laboratory frame.)

From Eq.~(\ref{asydm}) and $y=1/2$, we get  
$A_{RL}^{FT;ESM}(\mu e)=-8.216\times10^{-8}$ for $E_\mu=50$ GeV (or 
$s=0.05\,{\rm GeV}^2$) and $A_{RL}^{FT;ESM}(\mu e)=-3.122\times10^{-7}$ for 
$E_\mu=190$ GeV (or $s=0.19\,{\rm GeV}^2$). Taking into account the muon 
mass we get for the same conditions $A_{RL}^{FT;ESM}(\mu e)=
-5.917\times10^{-8}$ and $A_{RL}^{FT;ESM}(\mu e)=-2.913\times10^{-7}$, 
respectively.  Form Fig.~1 and the illustrative values given above we see  that for 
fixed target experiments the muon-mass effects can not be neglected. In fact, taking
into account the muon-mass effects in  $A_{RL}^{FT}$ will be crucial to correctly
interpret the experimental data. This is due to the high sensitivity of  $A_{RL}^{FT}$ 
with the electroweak mixing angle as it can be seen in Fig.~2. For instance, for 
$E_\mu =190$ GeV a 0.5\% change in the  $A_{RL}^{FT}$  value corresponds to 
a 0.04\% change in  $\sin^2(\theta_W)$, {\it i.e.}, a change from $0.2315$ to $0.2316$.

In the NA47 experiment the beam polarization was determined by 
measuring the cross-section asymmetry for the scattering of polarized muons 
on polarized atomic electrons. Since the elastic cross section for fixed target 
experiments is large ($\sim 4-11$) mb may be it is possible to extract information
about the $A_{RL}^{FT;ESM}(\mu e)$ asymmetry with unpolarized electrons. This
parameter is dominated by weak effects. 

For collider experiments, that would be feasible if the muon-electron
collider is constructed, we have that the asymmetry is large when comparing with the
corresponding for fixed target experiments. 
For instance $A_{RL}^{CO;ESM}(\mu e)=-0.024 $ 
for $E_\mu=190$ GeV ($\sqrt{s}\sim 380$ GeV) and $y=1/2$, as can be seen from Fig.~3. 
We show  only one curve in Fig.~3 because mass effects are not important in
high-energy colliders. 
We have also verified that  $A_{RL}^{CO;ESM}$ is as sensitive as  $A_{RL}^{FT;ESM}$ 
to the value of $\sin^2(\theta_W)$. However, for statistical purposes we have to observe
that the cross section for colliders is considerably smaller than that of fixed target
experiments: $\sigma^{CO} \sim 66$ nb for $\sqrt{s}\sim 380$ GeV.

Another interesting possibility is the case when both leptons are 
polarized. We can define an asymmetry $A_{R;RL}$ in which
one beam is always in the same polarization state, say right-handed, and 
the other  is either
right- or left-handed polarized (similarly we can define  $A_{L;LR}$):
 \begin{equation}
A_{R;RL}=\frac{d\sigma_{RR}-d\sigma_{RL}}{d\sigma_{RR}+d\sigma_{RL}},\qquad
A_{L;RL}=\frac{d\sigma_{LR}-d\sigma_{LL}}{d\sigma_{LL}+d\sigma_{LR}}.\qquad
\label{ar}
\end{equation}

In Fig.~4 we show the asymmetry  $A^{FT;ESM}_{R;RL}$ as a function of 
$y\equiv\sin^2(\theta/2)$ and we note that the muon--mass effects are relevant for a wide
range of $y$-values. 
For $y=1/2$ and $E_\mu=190$ GeV we have $A^{FT;ESM}_{R;RL}=0.2844$ while for the
same condition the purely QED contribution gives $A^{FT;QED}_{R;RL}=0.5689$. 
Hence, we see that like in the electron-electron M\o ller scattering
this asymmetry is sensitive to the neutral vector contributions. This is 
interesting since this asymmetry, as we said before, can already be measured 
in experiments like the NA47 at CERN showing that it could be useful for doing 
electroweak studies also. 
In fact, this experiment has 
already measured the asymmetry for polarized muon-electron scattering expected
in QED: the cross section asymmetry for antiparallel ($\uparrow\downarrow$)
and parallel ($\uparrow\uparrow$)~\cite{smc}. 
This corresponds in our notation to the asymmetry
\begin{equation}
\hat{A}_{RL}^{FT}=\frac{(d\sigma_{LR}+d\sigma_{RL})-(d\sigma_{LL}+d\sigma_{RR})}
{(d\sigma_{LR}+d\sigma_{RL})+(d\sigma_{LL}+d\sigma_{RR})}
\label{smc}
\end{equation} 
We have confirmed the
value obtained by the SMC Collaboration for this asymmetry {\it i.e.,} it ranges
from 0.01 to 0.05 for low values of $y$ ($\hat{A}_{RL}\approx y$).   
 $A_{R;RL}$   ($A_{L;RL}$) is not interesting for colliders experiments since it
 gives $\sim1$ ($\sim-1$) due to the fact that  $d\sigma_{RR}$ ($ d\sigma_{LL}$)
 largely dominates in Eq.~(\ref{ar})
                      
We have also investigated the effect of a doubly charged bilepton vector boson.
In some models with vector bileptons like $U^{--}$ the decay 
$\mu^-\to e^-e^-e^+$ (and similar ones) does occur. The branching ratio for 
this decay is $\sim10^{-12}$~\cite{pdg}.
This bound strongly limits the ${\cal K}_{\mu e}$ couplings. The branching 
fraction for $\mu\to 3e$ decay is
\begin{equation}
B(\mu\to 3e)\propto \left(\frac{{\cal K}_{\mu e}{\cal K}_{ee}}{M^2_U}\right)^2\,
\frac{1}{G_F^2}.
\label{mu3e}
\end{equation}
For the case $M_U=300$ GeV and ${\cal K}_{ee}\approx1$ the experimental value 
for the above branching ratio implies ${\cal K}_{\mu e}\sim10^{-6}$. 
Hence, we see that with such a large suppression the $U$-amplitudes given in
Eqs.~(\ref{ua}) are negligible. However, since we do not know if the $U$-vector
doubly charged bilepton (if it exists) prefers to couple in an almost diagonal 
way to charged leptons we are free to consider also the case 
${\cal K}_{ee}\approx10^{-6}$ and ${\cal K}_{\mu e}\sim1$. 
In the first case the $U$-contributions to cross section and asymmetry are 
negligible and the asymmetry is the same as in the ESM: 
$A^{CO;ESM}_{RL}(\mu e)=-0.024$ for $E_\mu=190$ GeV ($\sqrt{s}=380$ GeV). In the second case we have 
$A^{CO;ESM+U}_{RL}(\mu e)=-0.0092$ for the same experimental conditions. In the following we
will assume that we are in the first case and so we will not consider the
$U$-contribution to the asymmetry.

In the model there is also a $Z'$ neutral vector boson which couples with the
leptons as follows
\begin{equation}
{\cal L}_{NC}^{Z'}=-\frac{g}{2c_W}\,\left[\bar{l}_{aL}\gamma^\mu L_l{l}_{aL}
+\bar{l}_{aR}\gamma^\mu R_l{l}_{aR}+\bar{\nu}_{aL}\gamma^\mu L_\nu{\nu}_{aL}
\right]\,Z'_\mu,
\label{zpnc}
\end{equation}
with $L_l=L_\nu=-(1-4s^2_W)^{1/2}/\sqrt3$ and $R_l=2L_l$.
The $t$-channel $Z'$-exchange amplitudes are similar to those of the ESM $Z$ 
given in the Appendix A2 of Ref.~\cite{assi1} but now with the couplings $L$'s 
and $R$'s given in Eq.~(\ref{zpnc}). 

Taking into account the $Z'$ contributions  the  $A_{RL}(\mu e)$ asymmetry is
considerably enhanced as showed in Fig.~5.
We see from Fig.~5 that the measurement 
of the $A^{CO}_{RL}(\mu e)$ asymmetry will be appropriate for searching extra neutral 
vector bosons with a mass up to 1 TeV. Since in the 331 model the $Z'$ couplings with the 
leptons are flavor conserving we do not have additional suppression factors 
coming from mixing. Hence  the non-diagonal M\o ller scattering
($\mu e$) can be very helpful, even with the present experimental capabilities,
for looking for non-standard physics effects. We will not consider the $Z'$ 
contributions to the $A_{RL}$ and $A_{R;RL}$
asymmetries in fixed target experiments because they are very small.

Finally, we would like to stress that other related processes like
$\mu^+e^-\to\mu^-e^+$ and $e^-e^-\to\mu^-\mu^-(\tau^-\tau^-)$ deserve a detailed
study. In particular the $\mu^+e^-\to\mu^-e^+$ collision is interesting, as 
it was pointed out by Hou~\cite{hou}, because of its connection with 
$M-\overline{M}$
conversion ($M\equiv \mu^+e^-$ denote the muonium atom). 
The effect of resonance production involving lepton flavor violating couplings 
with hypothetical particles (like doubly charged scalar or vector bileptons, 
flavor changing neutral scalar bosons) at a $\mu e$ collider have been considered 
in Ref.~\cite{choi}. However, 
Barger {\it et al.,} have shown that the resonance production of the known 
particles puts severe limits on the respective cross section~\cite{barger}. 

\acknowledgments 
This research was partially supported by Conselho Nacional de 
Ci\^encia e Tecnologia (J.C.M. and V.P.) and fully financed  by Funda\c c\~ao 
de Amparo \`a Pesquisa do Estado de S\~ao Paulo (M. C. R.).

\appendix

\section{The amplitudes}
\label{sec:aa}
For a fixed target experiment $s\approx m^2_\mu+2m_eE_\mu$ where $E_\mu$ is the 
muon beam energy. In the center of mass frame $t=m^2_e+ m^2_\mu-2KE+
2[E^2(K^2-m^2_\mu)^{1/2}]\cos^2\theta$. ($E$ is the electron total energy and 
$K$ is the muon kinetic energy.)
Up to a $-e^2/q^2$ factor the photon amplitudes are
\begin{mathletters}
\label{af}
\begin{eqnarray}
M^\gamma_{RR;RR}(t)=M^\gamma_{LL;LL}(t)&=&4E\left[K\cos^2\frac{\theta}{2}+
(K-m_\mu)\left(
\frac{K+m_\mu}{K-m_\mu}\right)^{1/2}\left(\cos^2\frac{\theta}{2}+2
\sin^2\frac{\theta}{2}\right)\right],\nonumber \\ &\to& 
\quad 8EK\quad{\rm when}\quad m_\mu\to0;
\label{af1}
\end{eqnarray}

\begin{equation}
M^\gamma_{RR;RL}(t)=M_{LR;LL}=M^\gamma_{RL;RR}(t)=M^\gamma_{LL;LR}(t)
 =2m_\mu E\sin\theta\to0\quad {\rm when}\quad m_\mu\to0;
\label{af2}
\end{equation}
\begin{eqnarray}
M^\gamma_{RL;RL}(t)&=&
4E\left[K+(K-m_\mu)\left(\frac{K+m_\mu}{K-m_\mu}\right)^{1/2} \right]
\cos^2\frac{\theta}{2}\nonumber \\
&\to& \quad 8EK\cos^2\frac{\theta}{2}\quad {\rm when} \quad m_\mu\to0;
\label{af3}
\end{eqnarray}
\end{mathletters}

Up to a factor $-g^2/(q^2-M^2_Z)$ the $Z$ exchange amplitude is
\begin{mathletters}
\label{az}
\begin{eqnarray}
M^Z_{RR;RR}(t)&=&2(g_V+g_A)E\,\left\{(3g_A+g_V)K-(g_A-g_V)K
\cos\theta\right.\nonumber \\ &&\mbox{}+
\left.\left(\frac{K+m_\mu}{K-m_\mu}\right)^{1/2}\left[g_A K-m_\mu g_A 
+3g_VK-3m_\mu g_V+g_A K\cos\theta\right.\right.\nonumber \\ &&\mbox{}\left. 
\left.-  m_\mu g_A\cos\theta-g_V K\cos\theta+m_\mu g_V\cos\theta\right]
\right\},\nonumber 
\\ &\to & \quad 8KE(g_V+g_A)^2\quad{\rm when}\quad m_\mu\to0;
\label{az1}
\end{eqnarray}

\begin{eqnarray}
M^Z_{LL;LL}(t)&=&2(g_V+g_A)E\,\left\{(3g_A+g_V)K-(g_A-g_V)K
\cos\theta\right.\nonumber \\ &&\mbox{}+
\left.\left(\frac{K+m_\mu}{K-m_\mu}\right)^{1/2}\left[g_A K-m_\mu g_A 
+3g_VK-3m_\mu g_V+g_A K\cos\theta\right.\right.\nonumber \\ &&\mbox{}\left. 
\left.-  m_\mu g_A\cos\theta-g_V K\cos\theta+m_\mu g_V\cos\theta\right]
\right\},\nonumber 
\\ &\to & \quad 8KE(g_V-g_A)^2\quad{\rm when}\quad m_\mu\to0;
\label{az2}
\end{eqnarray}

\begin{equation}
M_{LR;LL}=M^Z_{RL;RR}(t)=M^Z_{LL;LR}=M^Z_{RR;RL}(t)=2m_\mu E(g^2_V-g^2_A)\sin\theta\to0\quad {\rm when}
\quad m_\mu\to0;
\label{az3}
\end{equation}
\begin{eqnarray}
M^Z_{RL;RL}(t)&=&4E(g^2_V-g^2_A)\left[ K+
(K-m_\mu) \left(\frac{K+m_\mu}{K-m_\mu}\right)^{1/2}
\right]\cos^2\frac{\theta}{2}\nonumber \\ &\to&
\quad 8EK(g^2_V-g^2_A)\cos^2\frac{\theta}{2},\quad {\rm when}\quad m_\mu\to0;
\label{az4}
\end{eqnarray}

\end{mathletters}
All other amplitudes vanish.

The $U^{--}_\mu$ vector bilepton contribution to the cross section are
small because of the suppression factor $\vert {\cal K}_{\mu e }\vert^2$ in the
amplitudes. The cross section of the $\mu e$ scattering is dominated by the
same fields of the ESM. However, in the $A_{RL}(\mu e)$ asymmetry 
this factor cancel out.

Up to a $ig^2\vert {\cal K}_{\mu e}\vert^2/2(s-M^2_U)$ factor the $s$-channel
$U$-exchange amplitudes are
\begin{mathletters}
\label{ua}
\begin{eqnarray}
M^U_{RR;RR}(s)&=&-M^U_{RR;LL}(s)=M^U_{LL;RR}(s)=M^U_{LL;LL}(s)
\nonumber \\ &=&2E(K-m_\mu)
\left[\left(\frac{K+m_\mu}{K-m_\mu}
\right)-1 \right]^2\sin^2\frac{\theta}{2}
\nonumber \\ &\to& 0\quad {\rm when} \quad m_\mu\to 0;
\label{um1}
\end{eqnarray}

\begin{eqnarray}
M^U_{RR;RL}(s)&=&M^U_{RR;LR}(s)=-M^U_{RL;RR}(s)=M^U_{RL;LL}(s)
\nonumber \\ &=&
-M^U_{LL;RL}(s)=M^U_{LL;LR}(s)=-M^U_{LR;RR}(s)=M^U_{LR;LL}(s)\nonumber \\ &=&2Em_\mu\sin\theta,
\label{ua2}
\end{eqnarray}
all these amplitudes go to zero when $m_\mu\to0$.

Finally,
\begin{eqnarray}
M^U_{RL;RL}(s)&=&-M^U_{RL;LR}(s)=-M^U_{LR;RL}(s)=M^U_{LR;LL}(s)
\nonumber \\ &=&
2E(K-m_\mu)\left[\left(\frac{K+m_\mu}{K-m_\mu}
\right)+1 \right]^2\sin^2\frac{\theta}{2}\nonumber \\
&\to&4K\,\sin^2\frac{\theta}{2}\quad {\rm when} \quad m_\mu\to0.
\label{ua3}
\end{eqnarray}
\end{mathletters}

\section{Differencial cross section for muon-electron FT experiment}
\label{sec:ab}
In the target frame
\begin{equation}
\frac{d\sigma}{d\Omega}=\frac{1}{64\pi^2m_e(E^2_\mu-m^2_\mu)^{1/2}}\,
\frac{\vert\vec{q}_2\vert^2}{(E_\mu+m_e)\vert\vec{q}_2\vert-(E^2_\mu-m^2_\mu)^{1/2}E'_\mu\cos\theta}
\;\vert M\vert^2,
\label{ab1}
\end{equation}
where
\begin{eqnarray}
\vert M\vert^2&=&4\left[4EK\cos^2\frac{\theta}{2}+4E(K-m_\mu)
\left(\frac{K+m_\mu}{K-m_\mu}\right)^{1/2}\cos\frac{\theta}{2} \right]^2
+16m_\mu^2\sin^2\theta E^2 \nonumber \\ &&\mbox{}
+128E(K-m_\mu)\left[EK\cos^2\frac{\theta}{2}+E(K-m_\mu) 
\left(\frac{K+m_\mu}{K-m_\mu}\right)^{1/2}\cos^2\frac{\theta}{2}\right]
\left(\frac{K+m_\mu}{K-m_\mu}\right)^{1/2}\sin^2\frac{\theta}{2}
\nonumber \\ &&\mbox{}+2\left[8E(K-m_\mu)\left(\frac{K+m_\mu}{K-m_\mu}
\right)^{1/2} \sin^2\frac{\theta}{2} \right]^2,
\label{ab2}
\end{eqnarray}
and $\vert\vec{q}_2\vert,   E'_\mu$ are the momentum and the energy in the target referential frame 
of the off going muon.

\narrowtext

\newpage

\begin{center}
{\bf Figure Captions}
\end{center}
\vskip .5cm
\noindent {\bf Fig.~1} The $A_{RL}(\mu e)$ asymmetry for the ESM in a fixed target experiment 
as a function of the muon energy with $y=1/2$ and  $\sin^2(\theta_W)=0.2315$ for  
$m_\mu=0$ (continuous line) and $m_\mu = 0.105$ GeV (dashed line).\\

\noindent {\bf Fig.~2} The $A_{RL}(\mu e)$ asymmetry for the ESM in a fixed target experiment 
with $E_\mu=190$ GeV  and $y=1/2$ as a function of $\sin^2(\theta_W)$.\\

\noindent {\bf Fig.~3} The $A_{RL}(\mu e)$ asymmetry  for the ESM
 in a collider experiment with $E_\mu=190$ GeV ($\sqrt{s}=380$ GeV) and $\sin^2(\theta_W)=0.2315$
as a function of $y$.\\

\noindent {\bf Fig.~4} The $A_{R;RL}(\mu e)$ asymmetry for the ESM 
 in a fixed target experiment with $E_\mu=190$ GeV and  $\sin^2(\theta_W)=0.2315$
as a function of $y$ for  $m_\mu=0$ (continuous line) and
$m_\mu = 0.105$ GeV (dashed line).\\

\noindent {\bf Fig.~5} The asymmetry $A_{RL}(\mu e)$ for the ESM ( dashed line)
and for the ESM plus the $Z'$ contribution (continuous line)
in a collider experiment with $E_\mu=190$ GeV ($\sqrt{s}=380$ GeV),
$\sin^2(\theta_W)=0.2315$, and $y=1/2$ as a function of $M_{Z'}$.  

%\noindent {\bf Fig.~3} The $A_{RL}(\mu e)$ asymmetry with the contribution
%of a $Z'$ vector boson in collider experiment with $E_\mu=190$ GeV
%as a function of $\sin^2(\theta_W)
%$.\\

%\noindent {\bf Fig.~4} The $A_{R;RL}(\mu e)$ asymmetry as a function of $y$.
%The photon contribution appears as a continuous line
%and the contribution of the ESM plus the $Z'$ contribution appears as a
%dashed line.  \\

%\noindent {\bf Fig.~5} The asymmetry defined in Eq.~(\ref{smc}) as a function of
%$y$ only the photon contributions. 
%\noindent {\bf Fig.~6} The asymmetry $A^{CO;ESM+Z'}_{RL}(\mu e)$ as a function 
%of $M_{Z'}$ for $E_\mu=190$ GeV.  

\end{document}